\documentclass[manuscript]{aastex6}
\usepackage{multirow}

\shorttitle{Stellar spectral classification}

\begin{document}


\title{SDSS-DR12 bulk stellar spectral classification: Artificial Neural Networks approach}


\author{S. Kheirdastan\altaffilmark{1}} \and \author{M. Bazarghan\altaffilmark{1}}
\affil{Department of Physics, Zanjan University, Zanjan 313, Iran}

\altaffiltext{1}{ Department of Physics, Zanjan University, Zanjan 313, Iran} 

\email{bazargan@znu.ac.ir}

\begin{abstract}
This 
 paper explores the application of Probabilistic Neural Network (PNN), Support Vector Machine (SVM) and k-means clustering as tools for automated classification of massive stellar spectra. The data set consists of a set of stellar spectra associated with the Sloan Digital Sky Survey (SDSS) SEGUE-1 and SEGUE-2, which consists of 400,000 data from 3850 to 8900 \AA~with 3646 data points each. We investigate the application of principal components analysis (PCA) to reducing the dimensionality of data set to 280, 400 and 700 components.We show that PNN can give fairly accurate spectral type classifications $\sigma_{RMS}=1.752$, $\sigma_{RMS}=1.538$ and $\sigma_{RMS}=1.391$ and K-means can classify these spectra with an accuracy of $\sigma_{RMS}=1.812$, $\sigma_{RMS}=1.731$ and $\sigma_{RMS}=1.654$ and SVM with the accuracy of $\sigma_{RMS}=1.795$, $\sigma_{RMS}=1.674$ and $\sigma_{RMS}=1.529$ across the 280, 400 and 700 components, respectively. By using K-means the classification of the spectra renders 38 major classes. Furthermore, by comparing the results we noticed that PNN is  more successful than K-means and SVM in automated classification.  \\
\end{abstract}





\section{Introduction}

Over the past decades, stellar population has increased because of large spectroscopic surveys such as RAdial Velocity Experiment (RAVE)  \citep{b1}  or the Sloan Digital Sky Survey (SDSS) \citep{b2}. Visual classification of flooded data stream by human experts is often subjective and needs extensive effort which is very time consuming. Automated classification using different statistical modeling can easily be implemented for very large numbers of these spectra.\\
Classification of stellar spectra requires a model which is based on stars information and detail of analysis. The MK classification of stellar spectra \citep{b3,b4} has long been an important tool in astrophysics which is still in use today. \\
A reliable stellar spectral classification pipeline is necessary to automatically exploit these data sets. Developing an automated data analysis tool is of great challenge with the coming future instruments and the astronomers show fair attention towards these techniques which gives a very fast and reliable way of data analysis. There are many kinds of techniques for the classification of astronomical data such as Support Vector Machine (SVM), Probabilistic Neural Networks (PNN), Self-Organizing Map (SOM), Expert System and K-means. We choose PNN which is a kind of multilayer neural network model and K-means clustering for their fast and efficient implementation and also SVM algorithm. \\ 
The efficiency of artificial neural networks in spectral classification is addressed in previous works such as; \citet{b5, b6,b7}, \citet{b8}, \citet{b9}, \citet{b10}, \citet{b11}, and \cite{b11-2}. Also the K-Means classification of spectra is used for different astrophysical contexts in these papers; \citet{b12,b13}, \citet{b14}, \citet{b15,b16}, and \cite {b17}. The performance of SVM can be found in \citet{b29}, \citet{b30} and \citet{b31}.  \\
We also show how the use of principal component analysis (PCA) can greatly compress the spectra and reduce the dimensionality of the data. Moreover, using PCA leads to a faster processing of ANN algorithm, as the dimension of data (spectrum) is reduced, and it reduces the complexity of the neural network and hence improves the classification accuracy.\\  This paper is organized as follows: Sects. 2 and 3 describe the data sets used in this study, their preprocessing and reduction procedure, Sect. 4 presents different classification methods implemented in this work, while Sect. 5 discusses our results.


\section{Data Sets and Pre-Processing}
\subsection{Data Selection}

The spectra come from SEGUE-2 \citep{b18} which is one of the four surveys (BOSS, SEGUE-2, APOGEE, and MARVELS) and SEGUE-1 of the SDSS III. SDSS has been in operation since 2000, using the 2.5 m telescope at Apache Point Observatory (APO) in the Sacramento Mountains in the southern New Mexico \citep{b19}. The initial surveys \citep{b20} carried our five broad bands filters (u, g, r, i, z) \citep{b21}. \\
In this paper, data sets are from the SDSS data release 10 (DR10) and data release 12 (DR12). 
The data sets of stellar spectra from DR10 and DR12 associated with SEGUE-1 and
SEGUE-2 with 100000 for training set and 300000 for test set are used. The members of training set are selected such that, it covers all the spectral types available in the downloaded data set. It is because, while training, the network must learn from the examples of all the spectral types. Hence, while testing the network can detect and recognize the given unknown spectra. We use catalogue Archive Server (CAS) to extract information of SEGUE-2 targets. The data set were just labeled as stars of classes of objects.

\subsection{Spectra Pre-Processing}

 We processed the spectra to have the same starting and ending wavelength scale at 3850 to 8900 \AA~. The spectra have units of flux per unit wavelength which contain 3646 data points after having the same wavelength scale. The spectra were then normalized to have equal intensities in order to meet the requirements of our algorithms. The details of data sets are summarized in figure \ref{1}. We have data array consisting of 100000 rows and, 3646 columns for train set and $300000\times3646$ data array as a test set.\\
In MK classification spectral types are shown by alpha numeric model. To evaluate the performance of our algorithms we dedicated a continuous scale of numbers 1-38 for them (Table \ref{1}).

\begin{figure}[h!]
\centering
 \includegraphics[width=1\columnwidth]{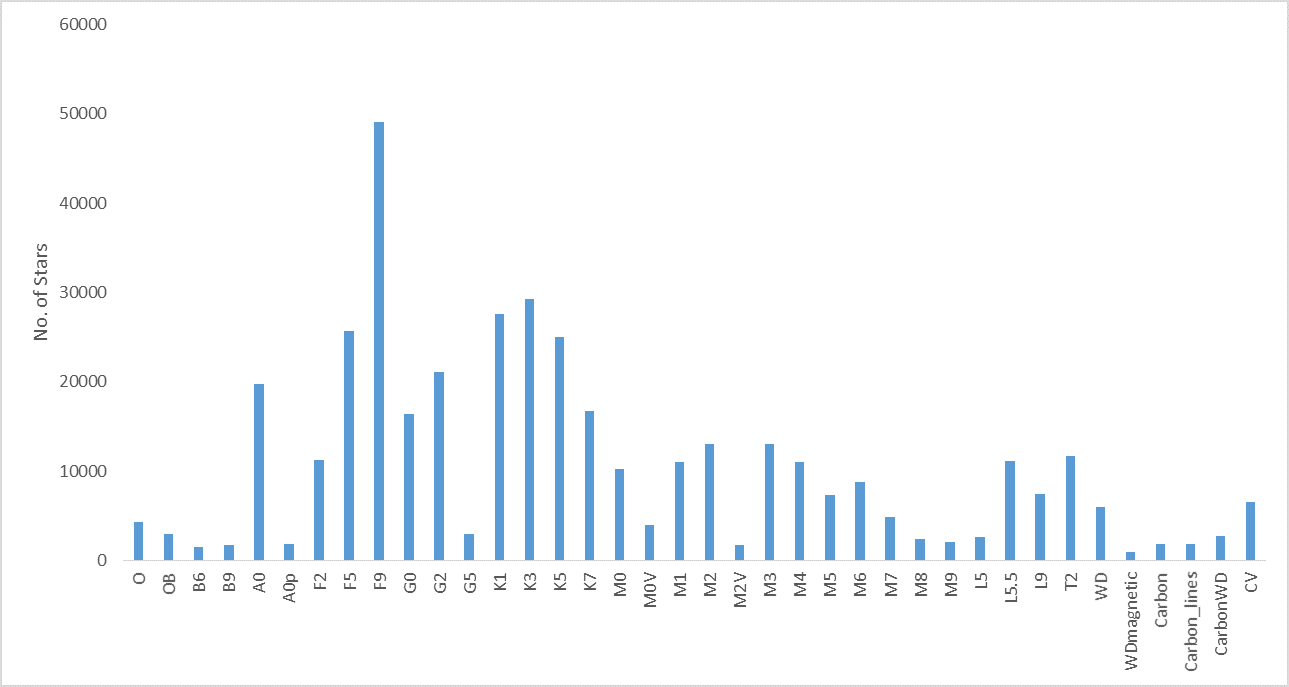}
 \caption{Population of stellar spectra types used from DR10 and DR12} 
 \label{1}
 \end{figure}

\begin{table}[h!]
\small
\begin{center}
\caption{Numerical coding used for each spectral type\label{tbl-2}}
 \label{1}
\begin{tabular}{@{}crrrrrrrrrr@{}}
\tableline
1 \,\,\,\,\,\,\,\,\,\,\,\,\,\,\,& O\,\,\,\,\,\,\,\,\,\,\,\,\,\,\,\,\,\,\,\,\,\,\,\,\,\,\,\,\,\,\,\,\,\,\,\,\,\,\,\,\,\,\,\,\,\,\,\,\,\,\,\, 20 \,\,\,\,\,\,&M2 \\
2\, \,\,\,\,\,\,\,\,\,\,\,\,\,\,\,& OB \,\,\,\,\,\,\,\,\,\,\,\,\,\,\,\,\,\,\,\,\,\,\,\,\,\,\,\,\,\,\,\,\,\,\,\,\,\,\,\,\,\,\,\,\,\,\,\,\, 21 \,\,\,\,\,\,&M2V \\
3\, \,\,\,\,\,\,\,\,\,\,\,\,\,\,\,& B6 \,\,\,\,\,\,\,\,\,\,\,\,\,\,\,\,\,\,\,\,\,\,\,\,\,\,\,\,\,\,\,\,\,\,\,\,\,\,\,\,\,\,\,\,\,\,\,\,\, 22 \,\,\,\,\,\,&M3 \\
4\, \,\,\,\,\,\,\,\,\,\,\,\,\,\,\,& B9 \,\,\,\,\,\,\,\,\,\,\,\,\,\,\,\,\,\,\,\,\,\,\,\,\,\,\,\,\,\,\,\,\,\,\,\,\,\,\,\,\,\,\,\,\,\,\,\,\, 23 \,\,\,\,\,\,&M4 \\
5\, \,\,\,\,\,\,\,\,\,\,\,\,\,\,\,& A0 \,\,\,\,\,\,\,\,\,\,\,\,\,\,\,\,\,\,\,\,\,\,\,\,\,\,\,\,\,\,\,\,\,\,\,\,\,\,\,\,\,\,\,\,\,\,\,\,\, 24 \,\,\,\,\,\,&M5 \\
6\, \,\,\,\,\,\,\,\,\,\,\,\,\,\,\,& A0p \,\,\,\,\,\,\,\,\,\,\,\,\,\,\,\,\,\,\,\,\,\,\,\,\,\,\,\,\,\,\,\,\,\,\,\,\,\,\,\,\,\,\,\,\,\,\, 25 \,\,\,\,\,\,&M6 \\
7\, \,\,\,\,\,\,\,\,\,\,\,\,\,\,\,& F2 \,\,\,\,\,\,\,\,\,\,\,\,\,\,\,\,\,\,\,\,\,\,\,\,\,\,\,\,\,\,\,\,\,\,\,\,\,\,\,\,\,\,\,\,\,\,\,\,\, 26\,\,\,\,\,\,\,\,&M7 \\
8\, \,\,\,\,\,\,\,\,\,\,\,\,\,\,\,& F5 \,\,\,\,\,\,\,\,\,\,\,\,\,\,\,\,\,\,\,\,\,\,\,\,\,\,\,\,\,\,\,\,\,\,\,\,\,\,\,\,\,\,\,\,\,\,\,\,\, 27\,\,\,\,\,\,\,\,&M8 \\
9\, \,\,\,\,\,\,\,\,\,\,\,\,\,\,\,& F9 \,\,\,\,\,\,\,\,\,\,\,\,\,\,\,\,\,\,\,\,\,\,\,\,\,\,\,\,\,\,\,\,\,\,\,\,\,\,\,\,\,\,\,\,\,\,\,\,\, 28\,\,\,\,\,\,\,\,&M9 \\
10\, \,\,\,\,\,\,\,\,\,\,\,\,\,\,\,& G0\,\,\,\,\,\,\,\,\,\,\,\,\,\,\,\,\,\,\,\,\,\,\,\,\,\,\,\,\,\,\,\,\,\,\,\,\,\,\,\,\,\,\,\,\,\,\,\,\,\,\, 29\,\,\,\,\,\,\,\,&L5 \\
11\, \,\,\,\,\,\,\,\,\,\,\,\,\,\,\,& G2 \,\,\,\,\,\,\,\,\,\,\,\,\,\,\,\,\,\,\,\,\,\,\,\,\,\,\,\,\,\,\,\,\,\,\,\,\,\,\,\,\,\,\,\,\,\,\,\,\, 30\,\,\,\,\,\,\,\,&L5.5 \\
12\, \,\,\,\,\,\,\,\,\,\,\,\,\,\,\,& G5 \,\,\,\,\,\,\,\,\,\,\,\,\,\,\,\,\,\,\,\,\,\,\,\,\,\,\,\,\,\,\,\,\,\,\,\,\,\,\,\,\,\,\,\,\,\,\,\,\,\, 31\,\,\,\,\,\,\,&L9 \\
13\, \,\,\,\,\,\,\,\,\,\,\,\,\,\,\,& K1 \,\,\,\,\,\,\,\,\,\,\,\,\,\,\,\,\,\,\,\,\,\,\,\,\,\,\,\,\,\,\,\,\,\,\,\,\,\,\,\,\,\,\,\,\,\,\,\,\, 32\,\,\,\,\,\,\,\,&T2 \\
14\, \,\,\,\,\,\,\,\,\,\,\,\,\,\,\,& K3 \,\,\,\,\,\,\,\,\,\,\,\,\,\,\,\,\,\,\,\,\,\,\,\,\,\,\,\,\,\,\,\,\,\,\,\,\,\,\,\,\,\,\,\,\,\,\,\,\, 33\,\,\,\,\,\,\,\,&WD \\
15\, \,\,\,\,\,\,\,\,\,\,\,\,\,\,\,& K5 \,\,\,\,\,\,\,\,\,\,\,\,\,\,\,\,\,\,\,\,\,\,\,\,\,\,\,\,\,\,\,\,\,\,\,\,\,\,\,\,\,\,\,\,\,\,\,\,\, 34\,\,\,\,\,\,\,\,&WDmagnetic \\
16\, \,\,\,\,\,\,\,\,\,\,\,\,\,\,\,& K7 \,\,\,\,\,\,\,\,\,\,\,\,\,\,\,\,\,\,\,\,\,\,\,\,\,\,\,\,\,\,\,\,\,\,\,\,\,\,\,\,\,\,\,\,\,\,\,\,\, 35\,\,\,\,\,\,\,\,&Carbon \\
17\, \,\,\,\,\,\,\,\,\,\,\,\,\,\,\,& M0 \,\,\,\,\,\,\,\,\,\,\,\,\,\,\,\,\,\,\,\,\,\,\,\,\,\,\,\,\,\,\,\,\,\,\,\,\,\,\,\,\,\,\,\,\,\,\,\,\, 36\,\,\,\,\,\,\,\,&Carbon-lines \\
18\, \,\,\,\,\,\,\,\,\,\,\,\,\,\,\,& M0V\,\,\,\,\,\,\,\,\,\,\,\,\,\,\,\,\,\,\,\,\,\,\,\,\,\,\,\,\,\,\,\,\,\,\,\,\,\,\,\,\,\,\,\,\,\,\,\,\, 37\,\,\,\,\,\,\,\,&CarbonWD \\
19\, \,\,\,\,\,\,\,\,\,\,\,\,\,\,\,& M1 \,\,\,\,\,\,\,\,\,\,\,\,\,\,\,\,\,\,\,\,\,\,\,\,\,\,\,\,\,\,\,\,\,\,\,\,\,\,\,\,\,\,\,\,\,\,\,\,\, 38\,\,\,\,\,\,\,\,&CV \\
\tableline
\end{tabular}
\end{center}
\end{table}

\subsection{Principal Component Analysis}

\begin{figure}[h!]
\centering
 \includegraphics[width=0.5\columnwidth]{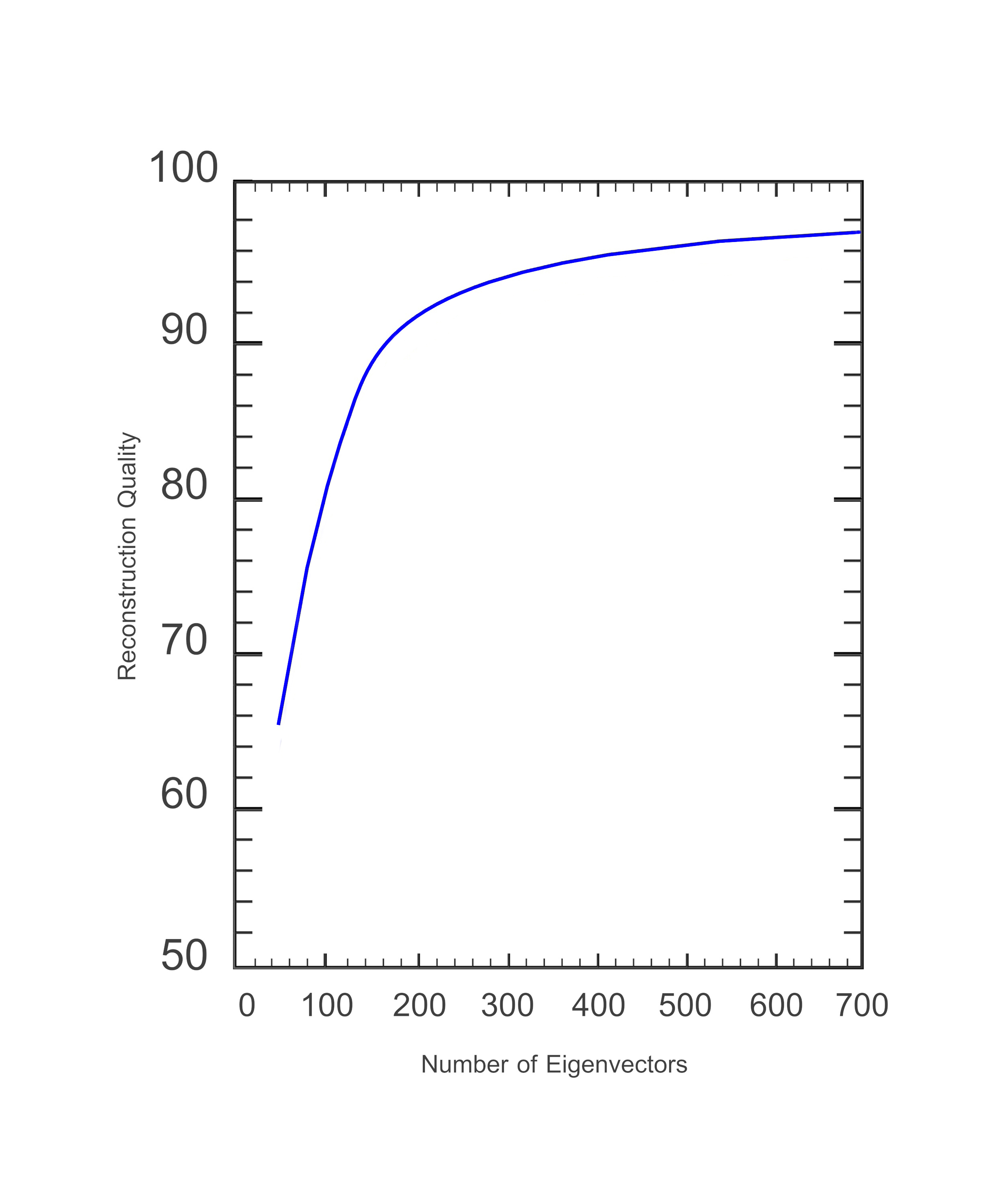}
 \caption{The curve showing the Quality of spectra reconstruction when the dimension of spectra is reducing: it grows from first 280 eigenvectors } 
 \label{7}
 \end{figure}

Principal Component Analysis (PCA) is a mathematical tool to emphasize variation and bring out strong patterns to reduce the dimensionality of data set. In particular, it allows us to identify the principal component direction in which the data varies. It often leads to make data set easy to classify. PCA has been used among others by \citet{b8} and \citet{b10} in stellar spectral classification, \citet{b22} for quasar spectral classification,  \citet{b23} and \citet{b24} for galaxy spectral classification.\\
The principal components are found by calculating the eigenvectors and eigenvalues of the data covariance matrix. This process is finding the axis system in which the covariance matrix is diagonal. The largest eigenvalues is the direction of greatest variation of eigenvectors, the one with the second largest eigenvalues is the orthogonal direction with the next highest variation and so on. Thus the most considerable principal component represent those features which vary the most between spectra ignoring those components that show the least variance in the data.\\
Wanting to get the spectra back is obviously of great concern in using  PCA transform for data compression but it is considerable that if we took all the eigenvectors in our transformation, we will get exactly the original spectra back. We reduced the number of eigenvectors in the final transformation, then the retrieved spectra lost some redundant data and noise. Detail of these analysis is described by \citet{b10} and we refer to that work. 
Reconstructed spectra contain the most significant principal components which are strongly correlated in many of the spectra. The quality of spectral reconstruction by increasing the number of eigenvectors is shown in figure \ref{7}. We see that the quality of reconstruction grow significantly over the first 280 eigenvectors that 280 eigenvectors is sufficient to reconstruct 94.6\% of the variance in the data and it rise to 97.2\% in 700 principal component. We use 280, 400 and 700 eigenvector to classify our data set to compare the result of our reduction. The advantage of this large dimensionality reduction includes providing a simpler representation of the spectra and fast classification resulted by the reduced complexity of the network and hence confusion of the system.

\section{Method of Classification}
\subsection{Probabilistic Neural Networks}

\begin{figure}[h!]
\centering
 \includegraphics[width=1\columnwidth]{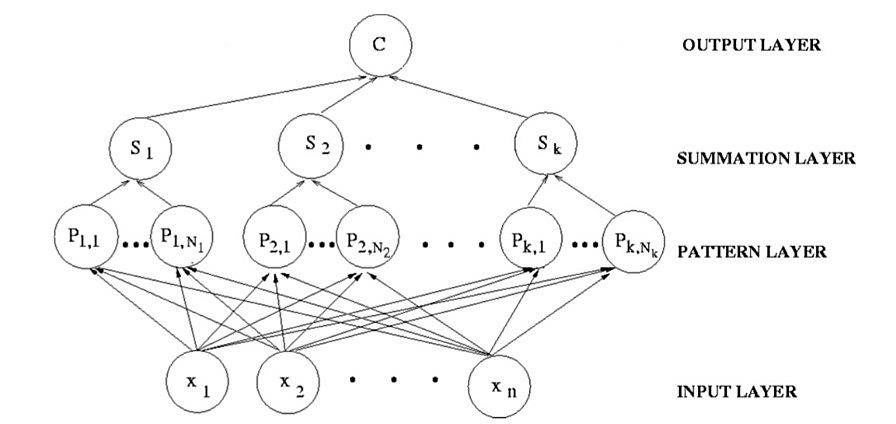}
 \caption{The architecture of Probabilistic Neural Network which consists of four layers } 
 \label{9}
 \end{figure}

The probabilistic neural network (PNN) described by \citet{b25}, is a special type of neural network based on Bayesian decision theory and the estimation of probability density function (PDF). PNN has a close relation with \citet{b26} window PDF estimator which is a nonparametric procedure. Unlike other neural networks, PNN is easy to implement and have fast training process. Because of these advantages, it has become an effective tool for solving classification problems. The PNN algorithm have been used extensively for the astronomical data \citet{b11}, \cite{b11-1}.  \\
The PNN architecture is composed of many interconnected neurons organized in four layer,  input layer, pattern layer, summation layer and output layer as illustrated in figure \ref{9}.\\

The input layer does not perform any computation and passes the input vector $X=(x_1,x_2,…,x_n )∈R^n$ to pattern layer. All connections in the network have a weight of 1, which means that the input vector is passed directly to each pattern node. In PNN, there is one pattern node for each training sample. Each pattern node has a center point, which is the input vector of sample. A pattern node also has a spread factor, which defines the size of its respective field. There are a diversity of ways to set this parameter. Spread factor is equal to the fraction f of the distance to the nearest neighbor of each sample. The value of f begins at 0.5. The neurons of the pattern layer compute its output using a Gaussian kernel,

\begin{equation}
S_{k,i}(x)=\frac{1}{(2\pi)^\frac{n}{2}\sigma^n}exp(-\frac{||X-X_{K,i}||^2}{2\sigma^2})
\end{equation}

where  n denotes dimension of the pattern vector $ X, \sigma$ is the smoothing parameter, and $X_{K,i}$ is the center of the Kernel. The summation layer neurons compute its output as the PDF for all neurons that belong to the same class.



Then at the output layer pattern $x$ is classified in accordance with the Bayes decision rule based on the output of all the summation layer neurons and the highest probability at the output of Summation layer will appear at the output layer c.

\begin{equation}
C(X) = arg\max_{1 \le k \le K}( S_k ),
\end{equation}

where  $K$ denotes the total number of classes in the training samples.

\subsection{K-means}

k-means was first used by  \citet{b27}. It is one of the simplest unsupervised learning algorithm that could solve the classification problems. The procedure follows a way to classify a given data set through a k number of clusters. Classification process begins by finding k number of clusters and their centers. Then each set of spectra is assigned to the nearest center that is closest in a least square sense. When all the spectra are loaded to the network, the cluster center gets refreshed (recalculated) as the average of the spectra in the cluster. And each point is replaced by the respective cluster center. These steps are repeated until no point moves cluster. Finally the algorithm provides a number of clusters with all the spectra assigned to one of them. Efficiency of K-means is closely related to the choice of  k number of clusters. We consider 38 clusters as the number of spectral type in our data set.

\subsection{Support Vector Machine}

Support Vector machine (SVM) is a supervised machine learning algorithm which was introduced by \citet{b28} and is based on the foundation of statistical learning theory. SVM is widely used for classification, data analysis and pattern recognition [\citet{b29}, \citet{b30} and \citet{b31}]. The main purpose of the SVM is to create a hyper plane or set of hyper planes in a high- or infinite-dimensional decision boundary between data sets to indicate which class it belongs to. The points on the boundaries are called support vectors. The problem of finding the optimal hyper plane is an optimization problem and can be solved by optimization techniques using Lagrange multipliers to get into a form that can be solved analytically. Whereas most of the other classifiers use all the data set to determine the boundary, SVM take the nearest points to the boundary. The challenge is to train the machine to provide the SVM with examples of the different classes of data set and dedicate right class label.  Training a SVM model belongs to a quadratic programming problem involving inequality constraints. The best result corresponds to the hyper plane that has the largest distance to the nearest training data point of any class (the maximum margin). Total classification error contain those points which lie on the wrong side of the hyper plane.


\section{Results and  Discussion}

\begin{figure}[h!]
\centering
 \includegraphics[width=0.5\columnwidth]{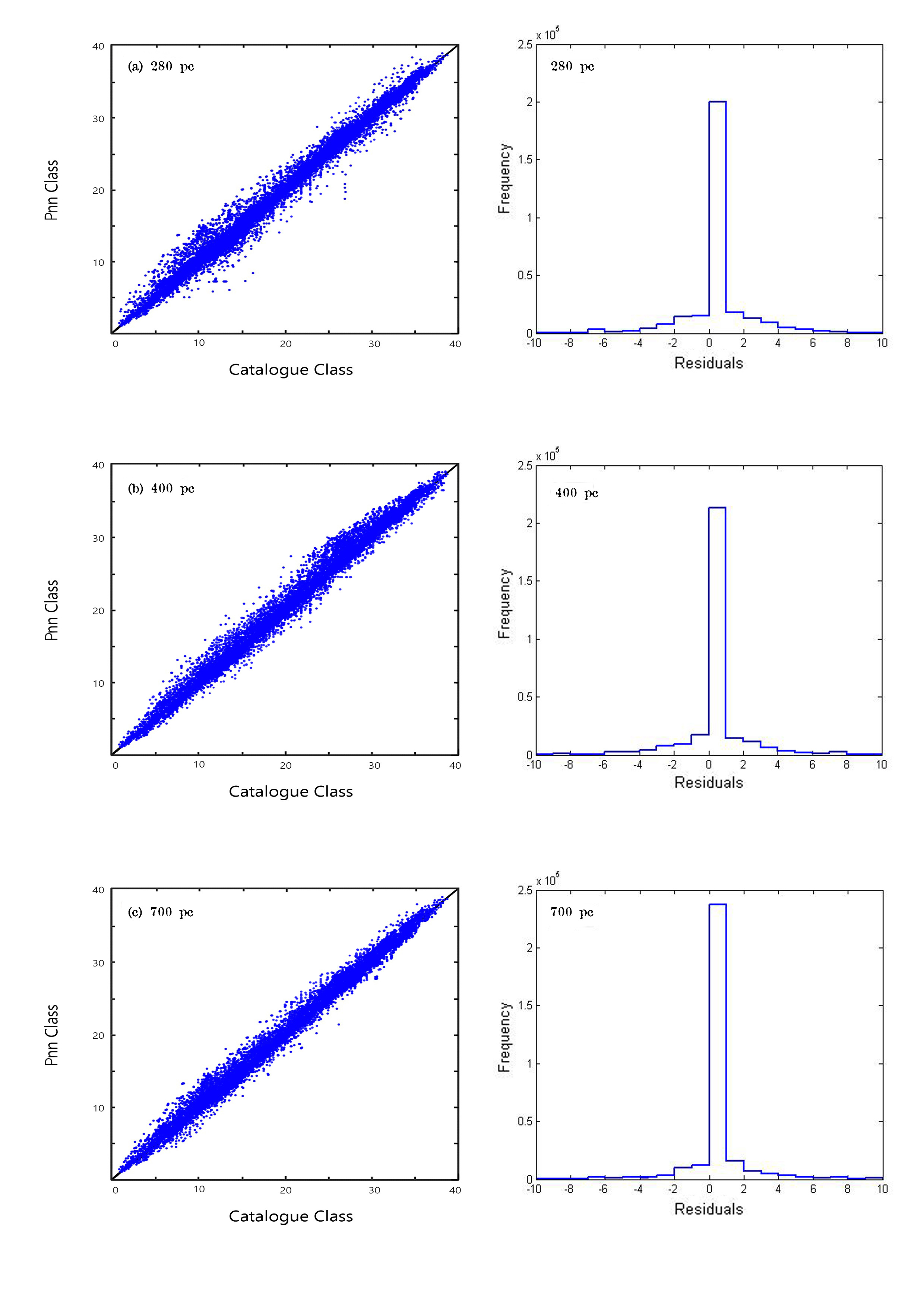}
 \caption{Spectral type classification results: left panels show the scatter plot and the high peak in the right panels show the correct classification rate for different principal components. The highest peak is for 700 PCs.} 
 \label{11}
 \end{figure}

In the following discussion we use PNN, SVM and K-means three times, for 280, 400 and 700 principal components. Figure 4 shows PNN classification results. In Figure \ref{11}a the classification scatter plot is shown, where 280 principal components of the spectra are used. In the subsequent panels result, Figure \ref{11}b and \ref{11}c, classification result of the 400 and 700 principal components are shown. The diagonal line plotted to better representation of the quality of PNN determination. Because of uncertainty in the classification, even an excellent classifier could not give results exactly on this line. \\

In Figure \ref{11}a it is clear that there is large scatter with the correlation coefficient of r=0.9314, thus indicating that 280 principal components are not at all sufficient for the satisfactory classification. Although configuration of 400 components gives a good result in this analysis with r=0.9328 but, the scattering was noticed to be better for 700 components with r=0.9356. There is small scatter in Figure 4c which convinced us that 700 principal components were enough to show the great quality of classification result. The right hand panels are histograms of the classification residual $(X^p-Y^p)$, where $X^p$  is the classification result and  the $Y^p$ catalogue class of spectrum. As we can see from the histograms, a total of 198121, 213540 and  237156 spectra have been classified correctly out of the total sample of 300000 spectra from 280, 400 and 700 principal components which indicating a success rate of approximately 66\%, 72\% and 80\%.

Considering on our best result (700 pc), we have the most misclassification for F9 stars. Out of a total 40637 spectra for this class, 31844 spectra were classified correctly. From the 8793 that were misclassified, most of them were classified as F5 and G0. F9 stars have been put in these classes by the PNN probably because of the resemblance of these spectra to each other and be the largest sample of our data set. The second numerous misclassification is for K3 which of a total of 22123 spectra, 15325 were correctly classified, while 6798 spectra were misclassified. The spectra were incorrectly classified into K1 and G2. M2 stars has 8844 spectra in all. 5376 spectra have been correctly classified, 3468 spectra have been misclassified into the other subclasses of M stars. PNN classification accuracy of WD has the best compliance with the class. Out of 2681 spectra of WD stars, 2653 spectra were classified correctly, while 28 spectrum was misclassified into other classes.\\

\begin{figure}[h!]
\centering
 \includegraphics[width=1\columnwidth]{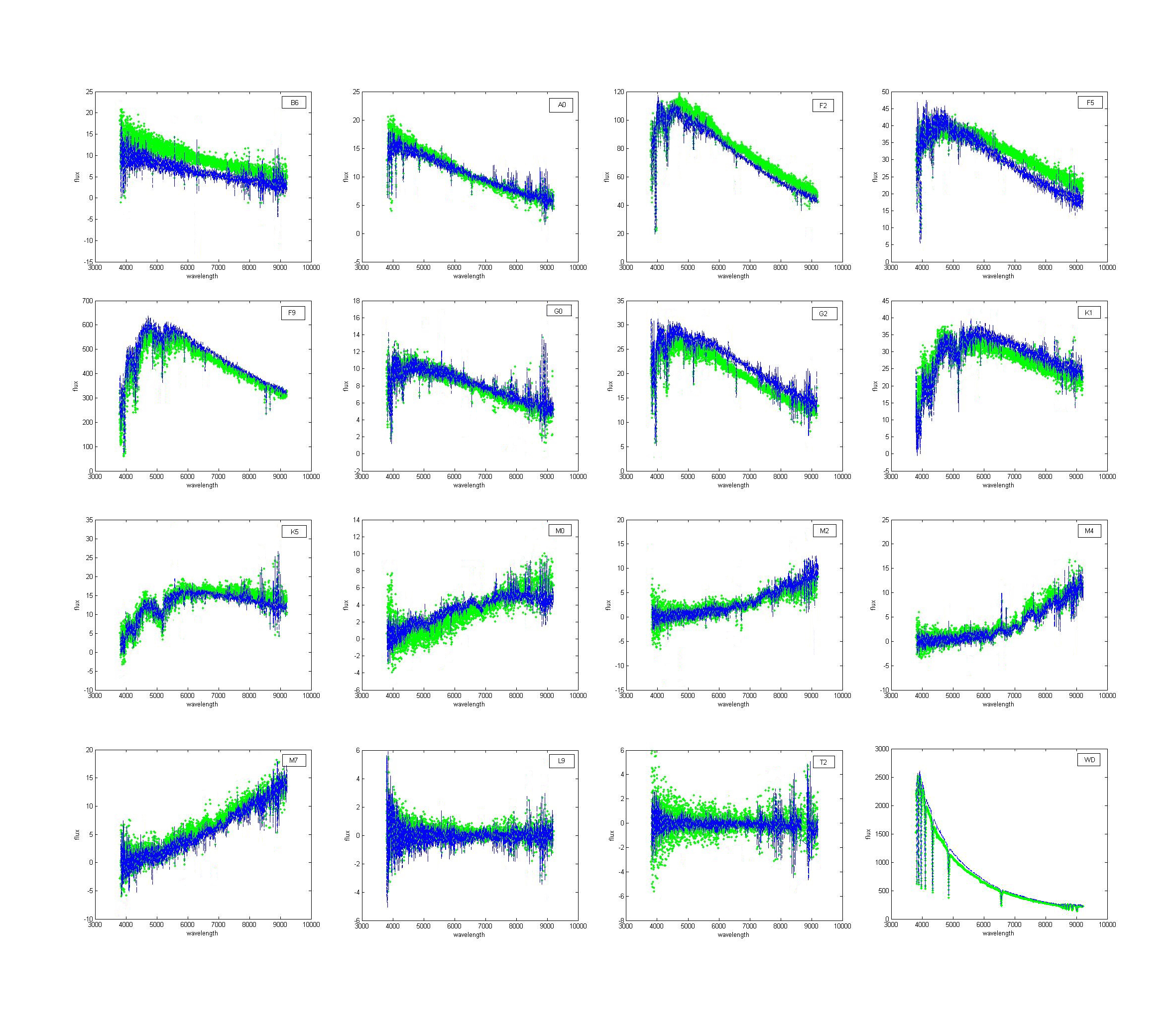}
 \caption{PNN classification results; pair spectra plotted in each window (result of classification Vs catalogue class) to show the goodness of fit. The catalogue spectrum is shown in dotted (green) line and the classification result in dashed (blue) line.} 
 \label{13}
 \end{figure}

Figure \ref{13} gives sample plots for correctly classified spectra from various spectral and sub-spectral types. It shows $4 \times 5$ windows with the pair spectra plotted in each window (result of classification Vs the catalogue class) to show the quality of classification and fitness. The catalogue spectrum is shown in dotted (green) line and the classification result in dashed (blue) line. \\

\begin{figure}[h!]
\centering
 \includegraphics[width=0.7\columnwidth]{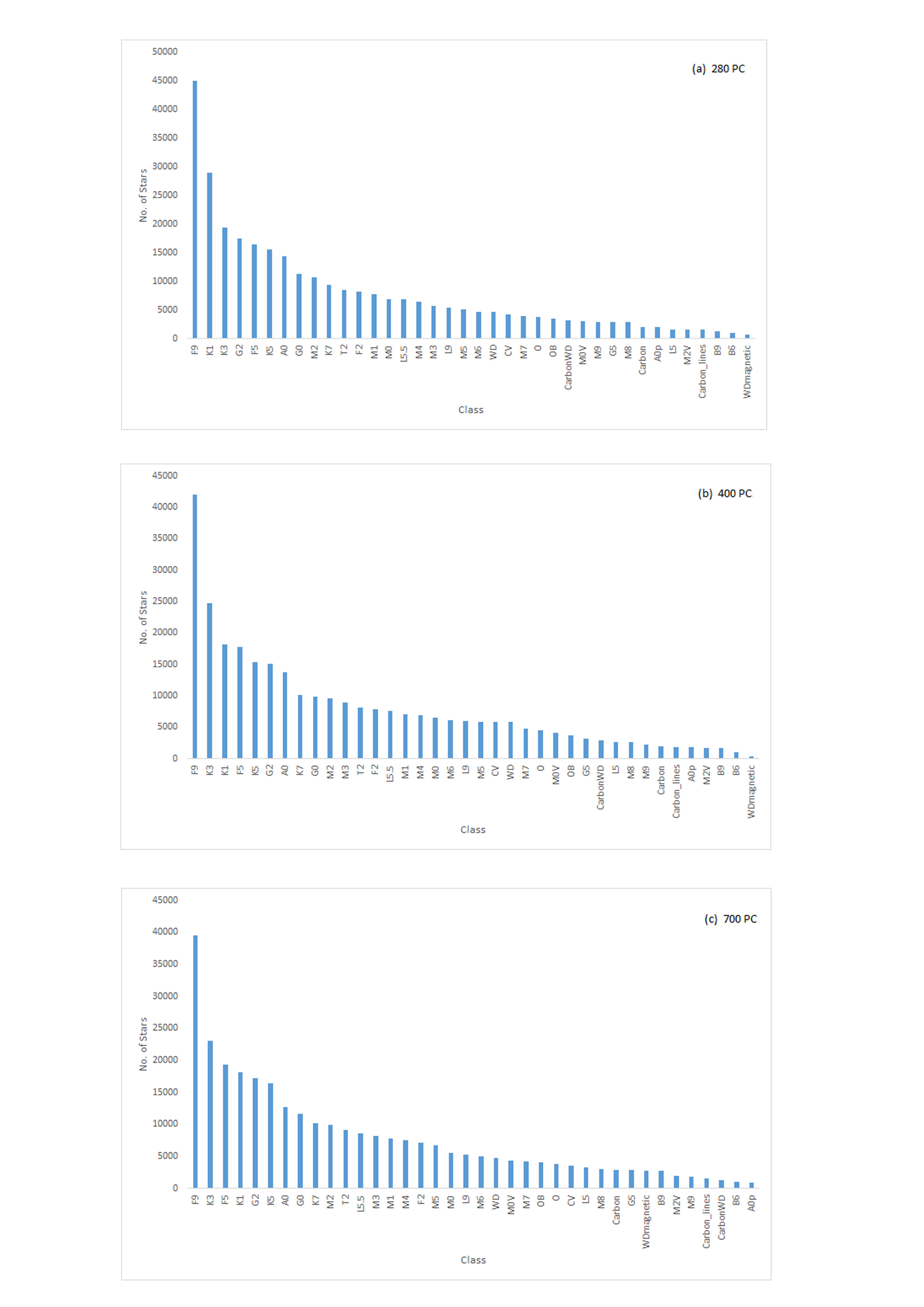}
 \caption{K-means classification results for different principal components. In result of classification using K-means, the population of each spectral type is plotted as histograms. } 
 \label{15}
 \end{figure}

The K-means algorithm is used for 300000 spectra with 280, 400 and 700 principal components to classify them into 38 clusters. The spectra in their classes are alike. The number of stars in the classes are shown as histograms in Figure \ref{15}, that Figure \ref{15}a, \ref{15}b and \ref{15}c belong to 280, 400 and 700 principal components respectively. These figures are arranged in the reducing order, the class with the highest population (highest peak) plotted first and the class with the lowest peak comes last. As a result of classification using K-means for the 280, 400, and 700 principal components, the highest peak belongs to the class of F9 stars. But the results show some variation for other classes using different components. \\

The allocation of the stars to each class depends on how far the star is from the cluster center. The error of this kind of classification could be measured by allocating each member of cluster to the right class and not similar to any other cluster. If most of stars in each cluster are the same as each other, the efficiency of classification increases. We obtained the best classification using 700 principal components, same as the classification with PNN. \\
Finally, we classified our data set with 280, 400 and 700 principal component by SVM algorithm. We use LIBSVM software which is provided by \citet{b32} to implement the SVM algorithm.There are four basic kernels in SVM: linear, polynomial, radial basis function (RBF), and sigmoid. When the number of features are very large, linear kernel is good enough and we just need to find an optimal C parameter, which is 0.7 in this project. Another important part that improves the result is, scaling before applying SVM. We normalize and use PCA on our data set which is discussed in section 2.2. After analyzing the SVM output we see that, some B type stars are misclassified as A type stars. Moreover, although the SVM classification works quite well for stars from O to A type, but F to K type stars have the highest completeness larger than 90\%. It means that more than 90\% F to K type stars are correctly classified by the SVM algorithm. The completeness of O to B and M type stars are about 82\% and 73\%, respectively, which are still acceptable. However, the completeness for other types of stars are only about 64\%. These error are probably because the spectral features of them are very similar and they are very difficult to be disentangled in SVM. Like PNN and K-means, the least error in SVM also belongs to 700 principal components.\\

Classification of spectra with 700 principal components by these algorithms needs more time than 280 and 400 principal components, but has better results. The most time consuming algorithm is SVM, then K-means and finally PNN has the least processing time.
The personal computer of core i7 with 12 cores CPU, 32 GB RAM and 256 GB SSD hard drive is used in this project. 
Table \ref{17} is a summary of the spectral classification results. We use RMS error to evaluate the performance of our algorithms. The most significant result is for the smallest error which is obtained using 700 principal components, as seen in the Table \ref{17} .  \\

\begin{table}[h!]
\caption{Error of classification and the processing time of different principal components using PNN, SVM, and K-Means algorithm}
\begin{center}
 \label{17}
    \begin{tabular}{ | l | l | l | l | l | l | l |}
    \hline
\multirow{2}{*}{ Principal Components} &
      \multicolumn{2}{c|}{280} &
      \multicolumn{2}{c|}{400} &
      \multicolumn{2}{c|}{700} \\
\cline{2-7}
   & $\sigma_{RMS}$  & Processing time& $\sigma_{RMS}$  & Processing time& $\sigma_{RMS}$  & Processing time\\
    \hline
    PNN & 1.752 & \hspace {1cm}{19 h} & 1.538 & \hspace {1cm}{20 h} & 1.391 & \hspace {1cm}{23 h} \\
    \hline
    SVM & 1.795 & \hspace {1cm}{22 h} & 1.674 & \hspace {1cm}{25 h} & 1.529 & \hspace {1cm}{29 h} \\
    \hline
K-means & 1.812 & \hspace {1cm}{21 h} & 1.731 & \hspace {1cm}{23 h} & 1.654 & \hspace {1cm}{26 h} \\
    \hline

  \end{tabular}
\end{center}
\end{table}


\section{Conclusion}
We have classified a set of 400000 stellar spectra with PNN, SVM and K-means algorithm. The PCA showed that the efficient dimension of the spectra is about 700 components. We have shown that we can achieve classification errors of 1.391, 1.529 and 1.715 for our best principal component (700 pc) using PNN, SVM and K-means respectively. These algorithms have been able to correctly classify approximately 80\% of the data set. As is summarized in Table \ref {17}, PNN and SVM have a better performance, i.e. lower RMS error, than the K-means algorithm for all feature sizes. PNN is marginally better than SVM for a relatively small feature size of 280 but as the feature size grows to 700, PNN shows a better performance than SVM.\\

\section{Acknowledgments}

We would like to thank Coryn Bailer Jones for his insightful suggestions.

\end{document}